\documentclass{ws-ijmpd}
\usepackage[super,compress]{cite}
\usepackage{bigints}
\usepackage{mathrsfs}
\usepackage{bm}
\newcommand{\gsim}{\raisebox{-0.13cm}{~\shortstack{$>$ \\[-0.07cm]
      $\sim$}}~}

\newcommand{\mdotin}{{\dot {\mathscr M}}_{\rm in}}
\newcommand{\rin}{r_{\rm in}}
\newcommand{\vin}{v_{\rm in}}
\newcommand{\vinp}{v_{\rm in}^\prime}
\newcommand{\rg}{r_{\rm g}}
\newcommand{\tg}{t_{\rm g}}
\newcommand{\tp}{T_{\mbox{{\scriptsize p}}}}
\newcommand{\te}{T_{\mbox{{\scriptsize e}}}}

\newcommand{\dqp}{\Delta Q_{\rm p}}
\newcommand{\dqe}{\Delta Q_{\rm e}}
\newcommand{\thetap}{\Theta_{\mbox{{\scriptsize p}}}}
\newcommand{\thetae}{\Theta_{\mbox{{\scriptsize e}}}}
\newcommand{\gamp}{\Gamma_{\mbox{{\scriptsize p}}}}
\newcommand{\game}{\Gamma_{\mbox{{\scriptsize e}}}}
\newcommand{\gamep}{Q_{\mbox{{\scriptsize ep}}}}
\newcommand{\polyp}{N_{\mbox{{\scriptsize p}}}}
\newcommand{\polye}{N_{\mbox{{\scriptsize e}}}}
\newcommand{\qpp}{Q^+_{\rm p}}
\newcommand{\qpm}{Q^-_{\rm p}}
\newcommand{\qep}{Q^+_{\rm e}}
\newcommand{\qem}{Q^-_{\rm e}}
\newcommand{\qbr}{Q_{\rm br}}
\newcommand{\qsy}{Q_{\rm syn}}
\newcommand{\qib}{Q_{\rm ib}}
\newcommand{\qic}{Q_{\rm ic}}
\newcommand{\gampc}{\Gamma_{\mbox{{\scriptsize p}}\scriptsize c}}
\newcommand{\gamec}{\Gamma_{\mbox{{\scriptsize e}}\scriptsize c}}

\newcommand{\mbh}{M_{\rm BH}}
\newcommand{\msol}{M_{\odot}}
\newcommand{\ep}{{\rm e}^--{\rm p}^+}

\newcommand{\nh}{n_{\rm in}}
\newcommand{\neh}{n_{\rm e in}}
\newcommand{\nph}{n_{\rm p in}}
\newcommand{\feh}{f_{\rm e in}}
\newcommand{\fph}{f_{\rm p in}}
\newcommand{\theh}{\Theta_{\rm e in}}
\newcommand{\thph}{\Theta_{\rm p in}}
\newcommand{\thehp}{\Theta_{\rm e in}^\prime}
\newcommand{\thphp}{\Theta_{\rm p in}^\prime}
\newcommand{\tph}{T_{\rm p in }}
\newcommand{\teh}{T_{\rm e in}}
\newcommand{\nel}{n_{\rm e}}
\newcommand{\np}{n_{\rm p}}
\newcommand{\me}{m_{\rm e}}
\newcommand{\mpr}{m_{\rm p}}
\begin{document}

\markboth{Shilpa Sarkar \& Indranil Chattopadhyay}
{Two-temperature accretion solutions around black hole}

%
\catchline{}{}{}{}{}
%

\title{General relativistic two-temperature accretion solutions for spherical flows around black holes
}

\author{Shilpa Sarkar 
}

\address{Aryabhatta Research Institute of Observational Sciences,\\ Manora Peak, Nainital, Uttarakhand 263002,
India
\\
shilpa@aries.res.in}

\author{Indranil Chattopadhyay}

\address{Aryabhatta Research Institute of Observational Sciences,\\ Manora Peak, Nainital, Uttarakhand 263002,
India
\\
indra@aries.res.in}

\maketitle

\begin{history}
\received{Day Month Year}
\revised{Day Month Year}
\end{history}

\begin{abstract}
Matter falling onto black holes is hot, fully ionized and has to be necessarily transonic. Since the electrons
are responsible for radiative cooling via processes like synchrotron, bremsstrahlung and inverse-Compton,
therefore the electron gas and proton gas are supposed to settle into two separate temperature distribution.
But the problem with two-temperature flow is that, there is one more variable than the number of equations.
Accretion flow in its simplest form is radial, which has two constants of motion.
While, the flow variables are,
the radial bulk three-velocity, electron and proton temperatures. Therefore, unlike single temperature flow, in the two temperature regime, there are multiple transonic solutions, non-unique for any given set of
constants of motion with a large variation in sonic points.
We invoked the second law of thermodynamics to find a possible way to break the degeneracy, by showing
only one of solutions among all possible, has maximum entropy and therefore is the correct solution.
By considering these correct solutions, we showed that the accretion efficiency increase with the increase in the
mass accretion rate. We showed that radial flow onto super-massive black hole can radiate with efficiency more
than 10\%, if the accretion rate is more than 60\% of the Eddington accretion rate, but accretion onto stellar-mass black hole achieve the same efficiency, when it is close to the Eddington limit. We also showed that, dissipative
heat quantitatively affects the two temperature solution. In presence of explicit heat processes the Coulomb coupling is weak.
\end{abstract}


\ccode{PACS numbers: 4.70.-s, 47.40.Hg, 51.30.+i, 95.30.Jx, 95.30.Lz, 97.10.Gz}



\section{Introduction}
One of the most spectacular objects found in the Universe are black holes (BH). Although not directly observable, their presence is interpreted from the huge amount of energy they liberate through a process called accretion. BH is an extreme compact object found, with sizes of the order of $\sim 3{\rm km}~(\mbh/\msol)$, where $\mbh$ is the mass of BH and $\msol$ is the solar mass. Due to their compactness, the amount of energy released due to accretion  might be of the order of the rest mass energy of the matter falling onto it. In the Universe, there exists stellar-mass BHs which accretes matter from a companion star and are visible in the sky as X-ray binaries, or may exist as super-massive ($\sim 10^6-10^9 \msol$) BHs which can feed on a full galaxy. Centres of such active galaxies are
famously known as the Active Galactic Nuclei (AGN) and are one of the brightest sources observed in the Universe. 

The advent of the theory of accretion onto compact objects began with the seminal works done by
Hoyle \& Lyttleton (1939) \cite{hoylelyt1939}  and Bondi (1952)\cite{bondi1952} , where they studied radial flow onto a gravitating centre. With the discovery of quasars \cite{schmidt63} and X-ray sources \cite{bowyer65} in 1960s, accretion of matter onto
compact objects gained popularity. That is because, accretion onto a BH is the only plausible mechanism which could
explain such high luminosities. In 1964, Salpeter \cite{salpeter1964} , computed the luminosity by using the Bondi accretion model, but failed to match it with observations. Matter being radially falling in case of Bondi flows, do not get sufficient time to radiate. The general relativistic version of Bondi flow was presented by Michel (1972)\cite{michel72} and this version of Bondi flow was also found out to be `too fast' to produce significant radiation \cite{shapiro73} . At about the same time, the famous Shakura-Sunyaev disc (SSD) or the Keplerian disc (KD) model was proposed \cite{shaksun1973} . Since Bondi flow 
could not explain the observed luminosity, therefore a rotation dominated disc model was envisaged
in order to mitigate the effect of very fast infall velocity. KD or SSD model assumes that matter is rotating 
with Keplerian azimuthal velocity, with an anomalous viscosity removing angular momentum outwards to accrete 
matter inwards. The heat generated is radiated away efficiently and the spectra produced by the disc is multi-coloured black body. 
Though this model could explain the thermal component of the spectrum but was unable to explain the high energy, non-thermal part of the spectrum. It was also realized that the flow should not be Keplerian everywhere especially near the BH. Therefore the inner region of an accretion flow has to be sub-Keplerian and has to pass through the sonic point
at least once before crossing the horizon as was shown by Liang \& Thompson \cite{liangthom1980} (hereafter, LT80). Subsequently, there was a significant body of work done by a number of workers on advective, transonic flows. Transonic flow has been studied for inviscid disc, viscous disc, around rotating BHs, discs harbouring shocks and host of other circumstances \cite{f87,nf89,c89,n92,mlc94,lmc98,ft04,fk07,cc11,kc13,kumaretal2013,kc14,ck16,kc17} . While for inviscid, adiabatic flow the sonic point can be obtained by solving a polynomial equation for a given
Bernoulli parameter and specific angular momentum, but obtaining sonic points for accretion flow in presence of heating and radiative cooling processes is not trivial. One can arbitrarily change the inner, or outer boundary conditions in order to obtain some solutions, but without a systematic approach it may lead to obtaining
limited class of solutions or a few unphysical ones. In that context, Gu \& Lu (2004)\cite{gulu04} used the generalized Bernoulli parameter which is also a constant of motion to obtain transonic viscous accretion solutions
for a particular viscosity prescription\cite{cm95} . 
The approach of Becker \& Le (2003)\cite{bl03} and Becker et. al. (2008)\cite{bdl08}, of using the generalized Bernoulli parameter simultaneously with the measure of entropy close to the horizon, to find the sonic point and therefore the transonic solution, is physically the most correct approach. Based on this approach, many papers
were written to obtain solutions in single temperature accretion flows around BHs\cite{kc13,kc14,ck16,lwwbp16,lb17} . Single temperature solutions are important to the extent that, it gives a general idea about the flow behaviour, its dynamics as well as energetics. To study the luminosity and spectra
of the accreting flow, one need to know electron temperature of the flow, which may or may not be same
as the proton temperature.

Due to the extreme gravity, matter falling onto a BH is very hot and becomes fully ionized. A fully ionized astrophysical plasma would be mostly composed of electrons and protons
because hydrogen is the most abundant element. Electrons radiate most of the energy
and protons do not, in addition, the Coulomb
coupling time scale is longer than the various cooling time scales, so in general,
protons and electrons would relax into two separate temperature distributions. In 1976, Shapiro Lightman \& Eardley \cite{shaplightear1976} (hereafter, SLE76)  argued that the instability persisting at the inner region of the disk could swell this optically thick radiation pressure dominated region into an optically thin gas pressure dominated region and in this region, electrons and protons will maintain separate temperature distributions. Since flow near the BH is a two-temperature fluid, as a result research on two-temperature accretion flow started to gain prominence \cite{svarts71,shaplightear1976} .

Two-temperature accretion solutions as presented by SLE76 incorporated inverse-Compton processes and could produce hard radiation. However, the hydrodynamics was significantly simplified, and the accretion solutions considered were not transonic. 
LT80 discussed primarily about single temperature transonic flow but also briefly discussed about two-temperature solutions by assuming the ratio of ion and electron temperature to be constant.
Since then, many studies were undertaken using two-temperature model. Colpi, Maraschi \& Treves (1984)\cite{colpi1984} solved the two-temperature solution but by assuming freely falling matter. No transonic solutions were reported here. Similar work was done by Chakrabarti \& Titarchuk (1995)\cite{ct95} where only inverse-Comptonization of soft photons from the SSD, by the inner post-shock region was considered.
Mandal \& Chakrabarti (2005)\cite{mandalchak2005} later extended this model for other cooling processes.
In both these papers, the authors imposed a density enhancement in the flow to mimic the accretion shock. In these works, the assumption of free-fall implied that the radiative transfer was not self consistently coupled with the hydrodynamics of the system. Laurent \& Titarchuk (1999)\cite{lt99} computed the spectra from the model of Chakrabarti \& Titarchuk (1995)\cite{ct95} , by conducting a detailed Monte-Carlo simulation of the interaction of electron gas with the soft photons from the underlying KD. In 1995, Narayan \& Yi (1995)\cite{narayanyi1995}, studied self-similar class of advective solutions (termed as advection dominated accretion flow or ADAF) in the two-temperature regime. It was assumed that the amount of heat transferred from protons to electrons through Coulomb collisions is totally radiated away. This extra assumption helped them to deal away with any kind of parametrisation, that was otherwise assumed by LT80. Needless to say self-similar class of solutions in conjunction with other assumptions mentioned above, are not transonic. Nakamura et al. (1996, 1997) \cite{nakaetal1996,nkmk97} , was among the first, who actually solved transonic two-temperature solution. However, the solutions were only for a limited class, obtained by imposing at the outer boundary, the ion temperature to be a fraction of the virial temperature and that the heat transferred to the electrons is radiated away. Manmoto et. al. (1997)\cite{manmotoetal97} followed similar outer boundary conditions,
however, in the inner region they considered that the electron energy advection rate to be equivalent
to the radiative cooling rate. Rajesh \& Mukhopadhyay
(2010)\cite{rajbani2010} also obtained transonic solutions in the two-temperature regime, by choosing the viscosity prescription of Chakrabarti \& Molteni (1995) \cite{cm95} , but only presented transonic solutions through a single sonic point. Dihingia et al.
(2017)\cite{induetal2017} discussed the transonic global two-temperature solutions for smooth as well as shocked accretion solutions.
All these works were in the pseudo-Newtonian regime (strong gravity is mimicked by modifying the
Newtonian gravitational potential), and used two fixed adiabatic indices ($\game$ \& $\gamp$ for electrons and
protons, respectively) equation of state (EoS) of the gas. None of these works used the constants of motion (e. g., generalized Bernoulli constant) of the flow and the information of entropy close to the horizon
to obtain the solutions. As discussed above, the hydrodynamics of single temperature regime is more
complete and systematic. In two temperature regime, this approach is sadly lacking in the literature.

The problem with two-temperature solutions is that, without any increase in the number of governing equations,
the number of flow variable increases, i.e. to say, now instead of a single temperature, one has to
consider different temperatures for ion and electron. In addition, there is no known principle dictated
by plasma physics which may constrain the relation between these two-temperatures in any of the boundaries.
Some authors (cited above) assumed specific relations between electron heating and cooling, in order to
obtain the solution. But those choices were arbitrary and cannot be considered a unique solution.
Such arbitrary choices do not `haunt' single temperature solution, since a transonic single-temperature solution
is unique for a given set of constants of motion.
Still some other authors followed the methodology of specifying the electron or ion temperature in a chosen boundary and then iterate the other flow variables to obtain a transonic solution.
However, a different combination of electron and ion temperature in that boundary can give rise to another
transonic solution but for the same value of generalized Bernoulli parameter \footnote{generalized Bernoulli parameter in steady state, is a constant of motion in presence of dissipation too \cite{kc13,kc14,ck16}}.
This would give rise to degeneracy of solutions, i. e., multiple transonic solutions for the same
set of constants of motion. But nature would prefer only one and the question is which one. Moreover, the electron and the ion temperature may vary by orders of magnitude from a large
distance to the horizon, so a non-relativistic EoS (i. e., EoS with fixed adiabatic indices) is untenable.
However, it may also be remembered that use of relativistic EoS even in single temperature domain has been traditionally few and far between \cite{bm76,f87,nf89} .

In this paper, we address the basic problem of finding a unique two-temperature transonic solution around BHs, in the general relativistic regime, using the two-temperature version of the Chattopadhyay-Ryu (CR)
EoS \cite{c08,cryu2009} , and how to overcome the problem, by laying down a prescription to obtain the correct
solution. As far as we know, such an attempt has not been undertaken before.
Using the CR EoS removes the constraint of specifying the adiabatic indices for the electron and the ion gas. In this paper, we would confine our discussion for a fully ionized electron-proton gas.     

The paper is organized in the following way. In Section 2, we present the assumptions and equations used in the paper which would cover the equation of state used and the equations of motion. In Section 3, we will discuss the procedure to obtain unique transonic two-temperature solutions. In Section 4, we will present and discuss our results and finally conclude in Section 5.

\section{ASSUMPTIONS AND EQUATIONS}

In this paper, we focus on obtaining the unique two-temperature solution in steady state from all the degenerate solutions,
which is actually difficult since plasma physics do not impose any constraint on the relation between
electron and proton temperatures at any distance from the BH horizon. Therefore, we remove all frills that might complicate and obscure the question at hand. As a first simplification, we consider Schwarzschild metric
i.e., the simplest BH. In order to further simplify the flow, we consider radial accretion i. e., rotation is neglected. Therefore the flow is spherical/conical. Although spherical accretion might look very simple,
however, it is not entirely implausible as an accretion model. In the viscous, single temperature, rotating accretion flow regime, we have previously shown that the flow geometry in the inner region of the disc is close to
conical flow with low angular momentum \cite{kc14,ck16} . Therefore, radial accretion might be used to
mimic the inner region of AGNs and microquasars. This is to be expected too, since the BH gravity would start to dominate
over other interactions in the inner accretion region around the BH, as a result a large number of papers do consider spherical flow to mimic the inner region of accretion flow \cite{ip82,tmk97,kht97} . In addition, standard accretion model onto
isolated BH from inter-stellar medium is indeed spherical\cite{dp80,bk05} .
We consider all possible
cooling mechanisms like bremsstrahlung, synchrotron
and inverse-Compton processes, and it may be noted that the electron is the main agent of emission. Energy is exchanged between electrons and protons through the Coulomb
interaction term given by Stepney (1983)\cite{stepney1983} . The effect of explicit heating is also discussed
at the end.

It is to be noted that in the subsequent sections, all barred variables represent dimensional quantities and all non-barred variables denote dimensionless quantities, until stated otherwise. Throughout this paper we have solved all the equations in the dimensionless domain. We have employed a unit system where, $G=\mbh=c=1$, such that the unit of length is $\rg=G\mbh/c^2$ and time is in units of $\tg=G\mbh/c^3$. Here, $G=$ Gravitational constant, $\mbh=$ mass of the BH and $c=$ speed of light.
\subsection{Equations of motion}
\label{sec:eom} 
The background metric is that around a Schwarzschild BH. The non-zero components of the Schwarzschild metric are,
\begin{equation}
\begin{split}
g_{tt}=-\left( 1-\frac{2}{r} \right) \textrm{ ;} \quad  g_{rr}={\left( 1-\frac{2}{r} \right)}^{-1} \textrm{ ;} 
~~~ g_{\theta \theta}=r^2 \textrm{ ;} \\ g_{\phi \phi}=r^2 \textrm{sin}^2 \theta,
	\label{eq:guv}
\end{split}
\end{equation}
The energy-momentum tensor of accretion flow is $T^{\mu \nu}=(e+p)u^\mu u^\nu + p g^{\mu \nu}$, where $e$
is the internal energy density of the fluid and $p$ is the isotropic gas pressure, all measured in local
fluid frame, $\mu$ and $\nu$ represent the space-time coordinates and $u^\mu$s are components of four velocities.
The space component of the relativistic momentum balance equation is given by,
\begin{equation}
    [(e+p)u^{\nu}u^i_{;\nu}+(g^{i\nu}+u^iu^\nu)p_{,\nu}]=0,
\label{eq:euler_1}
\end{equation}
The radial component of the above equation is given by,
\begin{equation}
    u^r \frac{du^r}{dr}+\frac{1}{r^2}=-(g^{rr}+u^r u^r) \frac{1}{e+p} \frac{dp}{dr},
	\label{eq:euler_1_radial}
\end{equation}
The equation of conservation of particle density flux is:
\begin{equation}
(nu^\nu)_{;\nu}=0,~\Rightarrow ~ \frac{1}{\sqrt{-g}}\frac{\partial (\sqrt{-g}nu^\nu)}{\partial x^\nu}=0.
	\label{eq:mass_cons}
\end{equation}
where, $n$ is the number density of the particles in the flow and $g$ is the determinant of the metric tensor.
Integrating equation(\ref{eq:mass_cons}), we get the accretion rate which is a constant of motion throughout the flow, given by,
\begin{equation}
    \dot{M}=4\pi \rho u^r r^2 cos(\theta),
	\label{eq:accretion_rate}
\end{equation}
where, $\theta$ is the co-latitude of the surface of the conical flow and is assumed to be
$\theta=60^{\circ}$ in this paper. The mass density is represented as $\rho$. 
The first law of thermodynamics is given by,
\begin{equation}
    u^\mu \left[ \left( \frac{e+p}{\rho} \right) \rho_{,\mu} -e_{,\mu} \right]=\Delta Q,
	\label{eq:flt_1}
\end{equation}
where, $\bm{\Delta Q=Q^+-Q^-}$, $\bm{Q^+}$ being the heating term and $\bm{Q^-}$ the cooling term. The dimensional form of any quantity are written with a bar over it, $\bm{\bar{Q}}$s are in units of ergs cm$^{-3}$ s$^{-1}$, until mentioned otherwise. The calculation of $\bm{\bar Q}$s require the value of number density (in units of cm$^{-3}$). The number density is calculated from the dimensional form of the accretion rate equation (in the dimensional form, the accretion rate is expressed in terms of Eddington rate). 

Since the flow contains electrons and protons equilibriating at two different temperatures we need to use the first law of thermodynamics separately for protons and electrons unlike in the case of one-temperature flows where the Coulomb coupling being extremely strong, allows protons and electrons to settle down to a single temperature \cite{kc13,kumaretal2013,kc14,ck16,kc17} . These two energy equations are coupled by the Coulomb coupling term which allows protons and electrons to exchange energy. 
Therefore, $\Delta Q$ in the proton energy equation can be written as, $\dqp=\qpp-\qpm$ and in the electron energy equation as, $\dqe=\qep-\qem$.

If we integrate the radial component of the relativistic Euler equation (\ref{eq:euler_1_radial}) with the help of equation (\ref{eq:accretion_rate}) and (\ref{eq:flt_1}), we obtain the generalized Bernoulli parameter which is a constant of motion and is given by,
\begin{equation}
    E=-hu_t \textrm{exp}(X_f),
	\label{eq:bernoulli_par}
\end{equation}
where,
$X_f=\int \frac{{\dqp}+{\dqe}}{\rho h u^r}dr$. This is conserved throughout the flow even in the presence of dissipation. In case of non-dissipative flows, $X_f=0$ and 
\begin{equation}
 E \rightarrow \mathcal E=-hu_t=h\gamma\sqrt{g_{tt}}
	\label{eq:bernoulli_par}
\end{equation}
where ${\cal {E}}$ is the canonical form of relativistic Bernoulli parameter \cite{cc11} for non-dissipative relativistic flow. The exact form of specific enthalpy $h$ will be presented in the next section
and $\gamma$ is the Lorentz factor.
%

%
%

\subsection{EoS and the final form of equations of motion}
\label{sec:eos} 
To solve the equations of motion mentioned in the previous section, we need an EoS which relates temperature, pressure and internal energy of the system. As discussed before we would use CR EoS \cite{cryu2009} . Since the adiabatic index is actually a function of temperature and composition, so it does not appear explicitly in the EoS. The CR EoS is inspired by the exact calculations done earlier \cite{c38,synge1957,cg68} . The
advantage of using CR over the exact EoS is that, the form of CR is much simpler and has been shown
to be equivalent \cite{vkmc15} .
The explicit form of CR EoS for multi-species flow is given by,
\begin{equation}
  {\bar {e}}=\sum_{i} {\bar {e}}_i=\sum_{i} \left[ \bar n_i  m_ic^2 +\bar p_i \left( \frac{9 \bar p_i+3 \bar n_im_ic^2}{3\bar p_i+2 \bar n_im_ic^2} \right) \right],
	\label{eq:eos}
\end{equation}
where, $i=\textrm {proton }(p)\textrm{, electron }(e^-) \textrm{, positron }(e^+)$ and $m_i$ is the mass of the corresponding i$^{\rm th}$ species.
In this paper, we consider the accretion flow to be electron-proton plasma ($\ep$).
So, in the sections to follow, $i$ would represent only protons and electrons. \\
We can define dimensional number density $(\bar n)$, corresponding mass density $(\bar\rho)$ and pressure $(\bar p)$ present in equation (\ref{eq:eos}) in the following way~:
\begin{equation}
  \bar n=\sum_{i} \bar n_i=\bar{n}_{\rm p}+\bar{n}_{\rm e}=2\bar{n}_{\rm e},
    	\label{eq:nd}
\end{equation}
where, $\bar{n}_{\rm p}=$proton number density, and  $\bar{n}_{\rm e}=$electron number density (in units of cm$^{-3}$).
\begin{equation}
 \bar \rho=\sum_{i} \bar n_i m_i= \bar{n}_{\rm e}\me+\bar{n}_{\rm p} \mpr=\bar{n}_{\rm e} \me \left( 1+\frac{1}{\eta}\right)=\bar{n}_{\rm e} \me \tilde{K},
	\label{eq:md}
\end{equation}
\begin{equation}
 \bar p=\sum_{i}  \bar p_i=\sum_{i} \bar n_ik T_i=\bar{n}_{\rm e} k(  {T}_{\rm e}+{T}_{\rm p})=\bar{n}_{\rm e} \me c^2\left(\thetae+\frac{\thetap}{\eta} \right),
	\label{eq:pressure}
\end{equation}
where, $\eta=\me /\mpr$, $\tilde{K}=1+1/\eta$, $T_i$ is the temperature of the $i^{\rm th}$ species (in units of Kelvin) and $k=$ Boltzmann constant. $\Theta_i=\frac{kT_i}{m_ic^2}$ is the non-dimensional temperature which has been defined w.r.t the rest-mass energy of the corresponding $i^{\rm th}$ species. \\
Using equations (\ref{eq:nd}) to (\ref{eq:pressure}) we can simplify the EoS (\ref{eq:eos}) to obtain,
\begin{equation}
{\bar e}=\bar{n}_{\rm e} \me c^2\left(f_{\rm e}+\frac{f_{\rm p}}{\eta} \right)=\frac{\bar \rho c^2f}{\tilde{K}},
	\label{eq:eos_sim}
\end{equation}
where, $f_i$ is defined as, $f_{i}=1+\Theta_i\left(\frac{9\Theta_i+3}{3\Theta_i+2}\right)$ and
$f=f_{\rm e}+f_{\rm p}/\eta$.

\noindent Enthalpy can be defined as,
\begin{equation}
{\bar h}=\frac{\bar e+\bar p}{\bar \rho}.
	\label{eq:enthalpy}
\end{equation}
Using equation (\ref{eq:md}), (\ref{eq:pressure}) and (\ref{eq:eos_sim}) we can reduce the above equation into a dimensionless form as,
\begin{equation}
    {h}=\frac{f+\left(\thetae+\thetap/\eta\right)}{\tilde{K}}.
	\label{eq:enthalpy_sim}
\end{equation}
The expression for polytropic index and adiabatic index for electrons and protons are given respectively as,
\begin{equation}
{N_{\rm p}}=\frac{df_{\rm p}}{d\Theta_{\rm p}};~ {N_{\rm e}}=\frac{df_{\rm e}}{d\Theta_{\rm e}} \textrm{;} \quad \Gamma_{\rm p}=1+\frac{1}{N_{\rm p}};~\mbox{and }\Gamma_{\rm e}=1+\frac{1}{N_{\rm e}}.
	\label{eq:poly_ad}
\end{equation}
The definition of the radial three-velocity is $v=[-(u_ru^r)/(u_tu^t)]^{1/2}$. Simplifying equations (\ref{eq:euler_1_radial}---\ref{eq:flt_1}, \ref{eq:eos}---\ref{eq:poly_ad}) we get the gradient of velocity,
\begin{equation}
    \frac{dv}{dr}=\frac{\mathcal{N}}{\mathcal{D}},
	\label{eq:dvdr}
\end{equation}
where, \\
$\mathcal{N}=-\frac{1}{r(r-2)}+{a^2 \mathfrak{P}}+(\game-1)\mathbb{E}+(\gamp-1)\mathbb{P} \quad$ and $\quad \mathcal{D}=\frac{v}{1-v^2}\left( 1- \frac{a^2}{v^2}\right)$.\\
Here, we have defined the sound speed as, $a^2=\mathcal{G}/h\tilde{K}$.\\
The expressions used in $\mathcal{N}$ and $\mathcal{D}$ are as follows,\\\\
$\mathcal{G}={\game\thetae}+\frac{\gamp \thetap}{\eta}$ ; $\quad\mathfrak{P}=\frac{2r-3}{r(r-2)}$ ;
$\quad \mathbb{E}=\frac{{\dqe}}{\rho h u^r}$ ; 
$\quad \mathbb{P}=\frac{{\dqp}}{\rho h u^r}$\\\\
Substituting equations (\ref{eq:accretion_rate}) , (\ref{eq:nd}) to (\ref{eq:eos_sim}) in equation (\ref{eq:flt_1}), we get the differential equation for both the proton and electron temperatures which is given by,
\begin{equation}
    \frac{d\thetap}{dr}=-\frac{\thetap}{\polyp}\left(\mathfrak{P}+ \frac{1}{v(1-v^2)}\frac{dv}{dr}\right) - \frac{\mathbb{P}\eta \tilde{K}h}{\polyp},
	\label{eq:dtpdr}
\end{equation}
\begin{equation}
    \frac{d\thetae}{dr}=-\frac{\thetae}{\polye}\left(\mathfrak{P}+ \frac{1}{v(1-v^2)}\frac{dv}{dr}\right) - \frac{\mathbb{E}\tilde{K}h}{\polye},
	\label{eq:dtedr}
\end{equation}
respectively.
%
\subsubsection{Radiative processes considered}
\label{sec:rt} 
 Cooling of protons can be caused due to Coulomb interactions ($\bm{\bar{Q}_{\rm{ep}}}$) with electrons
if $\tp > \te$, or due to inverse bremsstrahlung ($\bm{\bar{Q}_{\rm{ib}}}$). 
The expression for Coulomb interaction term in cgs unit is given by \cite{sg83} ,
\begin{align}
      & \bar{Q}_{\rm{ep}} = \frac{3}{2}\frac{\me}{\mpr}\bar{n}_{\rm e}\bar{n}_{\rm p}  \sigma_T c k \frac{\tp-\te}{K_2 \left(1/\thetae\right) K_2 \left( {1/\thetap}\right)} \textrm{ln } \Lambda_c \nonumber \\ 
         & \times \left[ \frac{2(\thetae+\thetap)^2+1}{\thetae+\thetap} K_1 \left( \frac{\thetae+\thetap}{\thetae \thetap}\right)+ 2K_0 \left( \frac{\thetae+\thetap}{\thetae\thetap}\right)  \right],
\label{eq:cc}
\end{align}
where, $\sigma_T$ is the Thomson scattering cross-section, $K_i(x)$'s are the modified Bessel functions of i$^{\rm th}$ order and second kind, ln $\Lambda_c$ is the Coulomb logarithm which we took to be equal to 20.\\
The expression for inverse bremsstrahlung is given by \cite{bs69,j71} ,
\begin{equation}
\bar{Q}_{\rm{ib}} =1.4 \times 10^{-27} \bar{n}_{\rm e}^2 \sqrt{\frac{\me}{\mpr}\tp}.
	\label{eq:ib}
\end{equation}

The cooling of electrons includes contributions from three radiative cooling mechanisms namely bremsstrahlung ($  \bar{Q}_{\rm{br}}$), synchrotron ($\bar{Q}_{\rm{syn}}$) and inverse Compton scattering ($  \bar{Q}_{\rm{ic}} $). Therefore, $\bar{Q}_{\rm{e}}^- =\bar{Q}_{\rm{br}}+\bar{Q}_{\rm{syn}}+\bar{Q}_{\rm{ic}}$.\\
The expression for bremsstrahlung emissivity (in c.g.s units) is given by \cite{nt73},
\begin{equation}
\bar{Q}_{\rm{br}}=1.4 \times 10^{-27} \bar{n}_{\rm e}^2 \sqrt{\te}(1+4.4 \times 10^{-10}\te).
	\label{eq:brem}
\end{equation}
The cooling per unit volume in case of synchrotron radiation is given as \cite{narayanyi1995} ,
\begin{equation}
 \bar{Q}_{\rm{syn}}=\frac{2\pi k \te}{3c^2} \frac{\nu_c^3}{r\rg},
	\label{eq:sync}
\end{equation}
where, $\nu_c$ is the critical frequency below which the emission is self-absorbed. It can be defined as $\nu_c=\frac{3}{2}\nu_o \thetae^2 x_M$ where $\nu_o=2.8 \times 10^6 B$. One has to solve a transcendental equation
to obtain the value of $x_M$. Here, $B$ is defined as the stochastic magnetic field present in the flow, whose value is obtained by assuming its pressure ($B^2/8\pi$) to be in partial or full equipartition with the gas pressure ($\bm{\bar p}$). This ratio can be defined as $\beta$ and is chosen as $\beta = 0.01$, unless otherwise mentioned.\\
The Comptonization of the soft photons generated through synchrotron process, is given as
\cite{dermer1991} ,
\begin{equation}
\bar{Q}_{\rm{ic}}={\zeta}\bar{Q}_{\rm{syn}},
	\label{eq:comp}
\end{equation}
where, $\zeta$ is the enhancement factor which is defined as the average change in energy of the photon at escape after all scatterings. It is expressed as $\zeta=P(A-1)(1-PA)^{-1}\left[ 1- \left({x_c/(3\thetae)}\right)^{-\left(1+{\textrm{ln}P/\textrm{ln}A}\right)}\right]$. Here, $x_c=h\nu_c/\me c^2$, $P=1-\textrm{exp}(-\tau_{\textrm{es}})$ is the probability of a photon to be scattered in a medium with optical depth $\tau_{\textrm{es}}$, and $A=1+4\thetae+16\thetae^2$, is the mean amplification factor in energy of the scattered photon. The optical depth of a medium where electron-scattering is dominant is given by\cite{turolla1986} , $\tau_\textrm{es}=0.4 \left[ 1+ \left(2.22 \te \times 10^{-9} \right) ^{0.86}\right]^{-1}$.

The plasma is heated via magnetic dissipation and it primarily affects the proton distribution, and part of this
heat is transmitted to the electrons through the Coulomb coupling term. The dissipative heating rate is given
by\cite{ip82,bk05} ,
\begin{equation}
\bar{Q}_{\rm{p}}^+ \approx {\bar{Q}}_{\rm B}=\frac{3cu^r}{2r r_g}\frac{B^2}{8\pi} = \frac{3cu^r}{2r r_g} \beta \bar{n}_{\rm e} k (\te + \tp).
\label{eq:dissiphit} 
\end{equation}
 
\subsubsection{Entropy accretion rate expression}
\label{subsec:entropy_accretion_formula}
From single temperature solutions we know we can define an entropy-accretion rate by integrating equations
(\ref{eq:dtpdr} \& \ref{eq:dtedr}) by turning off the explicit heating and cooling terms.
\begin{align}
\frac{d\thetap}{dr}=\frac{\thetap}{\polyp}\frac{1}{\np}\frac{d\np}{dr} + \frac{\gamep \eta \tilde{K}}{\rho u^r \polyp} \nonumber \\ 
 \frac{d\thetae}{dr}=\frac{\thetae}{\polye}\frac{1}{\nel}\frac{d\nel}{dr} -
 \frac{\gamep\tilde{K}}{\rho u^r \polye}.
\label{eq:fst1law2}
\end{align}
In single temperature regime, it is very easy to integrate the above equation, but now, due to the
presence of Coulomb interaction term, equation (\ref{eq:fst1law2}) is not generally integrable. And therefore,
we cannot have an analytical expression for the measure of entropy at every $r$ in two-temperature solutions.

However, in regions where $\gamep$ can be neglected, an analytical expression is admissible. Such a region is
just outside the horizon, where gravity overwhelms any other interaction. 
So, near the horizon, where $\gamep$ is negligible, equations (\ref{eq:fst1law2}) can be integrated to obtain,
\begin{align}
&\neh=\kappa_1 ~ {\rm exp}{\left({\frac{\feh-1}{\theh}}\right)}\theh^{\frac{3}{2}}(3\theh+2)^{\frac{3}{2}}\\
&\nph=\kappa_2 ~{\rm exp}{\left({\frac{\fph-1}{\thph}}\right)}\thph^{\frac{3}{2}}(3\thph+2)^{\frac{3}{2}},
\label{eq:ent_2}
\end{align}
where, $\kappa_1$ and $\kappa_2$ are the integration constants which are a measure of entropy.
We know from charge neutrality condition that $\neh=\nph=\nh$. Subscript `${\rm in}$' indicates quantities measured just outside the horizon. Therefore we can write,
\begin{align}
\nonumber \nh^2 &=\neh \nph\\ \nonumber
\Rightarrow \nh &=\sqrt{\neh \nph}\\ \nonumber
&=\kappa \sqrt{{\rm exp}{\left({\frac{\feh-1}{\theh}}\right)}~{\rm exp}{\left({\frac{\fph-1}{\thph}}\right)}\theh^{\frac{3}{2}}\thph^{\frac{3}{2}}} \\
& \times \sqrt{(3\theh+2)^{\frac{3}{2}}(3\thph+2)^{\frac{3}{2}}},
\end{align}
where, $\kappa=\sqrt{\kappa_1 \kappa_2}$\\
Thus, the expression of entropy accretion rate can be written as,
\begin{align}
\nonumber \mdotin &=\frac{\dot{M}}{4\pi\kappa (\me+\mpr){\rm cos}(\theta)}\\ \nonumber
&=\sqrt{{\rm exp}{\left({\frac{\feh-1}{\theh}}\right)}{\rm exp}{\left({\frac{\fph-1}{\thph}}\right)}\theh^{\frac{3}{2}}\thph^{\frac{3}{2}}} \\
& \times \sqrt{\left((3\theh+2)^{\frac{3}{2}}(3\thph+2)^{\frac{3}{2}}\right)} u^r r ^2
\label{eq:ent_acc_rate}
\end{align}

In section \ref{subsec:uniqsoln}, we will use equation (\ref{eq:ent_acc_rate}) to obtain the correct accretion solution. 

\subsubsection{Sonic point conditions}
\label{subsec:sonic_point_cond} 
As argued before, black hole accretion is transonic in nature. So, at some $r=r_c$ the critical point, the flow $dv/dr \rightarrow 0/0$. This condition gives us the critical point conditions. Thus using equation (\ref{eq:dvdr}) we get,
 \begin{align}
-\frac{1}{r_c(r_c-2)}+{a_c^2 \mathfrak{P}_c}+(\gamec-1)\mathbb{E}_c+(\gampc-1)\mathbb{P}_c=0,
	\label{eq:sp1}
\end{align}
and,
\begin{equation}
 \frac{v_c}{1-v_c^2}\left( 1- \frac{a_c^2}{v_c^2}\right)=0.
	\label{eq:sp2}
\end{equation}
Here, `$c$' in the subscript resembles the values of the variables at the critical point.
At $r_c$, the radial three-velocity is equal to the sound speed or $v_c=a_c$, i. e., the Mach number $M_c=v_c/a_c=1$.
Since the derivative of velocity at the critical point has a $0/0$ form, therefore it is calculated using l$'$Hospital rule.
%
%

\section{SOLUTION PROCEDURE}
\label{sec:solution}
It has already been established by Bondi (1952)\cite{bondi1952} , that for a given boundary condition the entropy of the
transonic global solution is maximum, and therefore, a transonic solution is the solution favoured by nature. Therefore, we look for a transonic solution. The general procedure to find a solution in two-temperature is similar to the one in the single temperature regime, which is --- for a given set of flow parameters ($E$, $\dot{M}$), the sonic point is obtained first, and then integrate the gradient of velocity and temperature, that is equations (\ref{eq:dvdr}), (\ref{eq:dtpdr}) and (\ref{eq:dtedr}), from the sonic point inwards and outwards, in order to obtain self-consistent values of $v$, $\thetap$ and $\thetae$ respectively throughout the flow. A spherical flow harbours only a single sonic point. 
 
\subsection{Method to find the sonic point: single temperature versus two temperature}
\label{subsec:method1}
This is the first step in obtaining a general transonic solution. Finding a sonic point is not trivial in presence of heating and cooling. To find the sonic points we need to first choose a boundary: horizon or infinity. 
The advantage of choosing the horizon as the boundary, is that atleast the inflow velocity on the horizon is known ($\vin=c$), while at the outer boundary its value is arbitrary. Unfortunately, there is a coordinate singularity on the horizon, so one cannot start the integration from the horizon. Therefore, we chose a location asymptotically very close to the horizon, $\rin \rightarrow 2\rg$. Very close to the horizon gravity overwhelms all other interactions, therefore the flow becomes adiabatic, i. e., as $\rin \rightarrow 2 \rg$,  $E \rightarrow {\cal E}$.
At $\rin$ for single temperature flow, there are two unknowns $\vin$ and the temperature. So for a given $E$, at $\rin$ we supply a temperature in the expression of $E={\cal E}$ to obtain a value of velocity, say $v_{\rm in}^\prime$. With these values of velocity and temperature we integrate the equations of gradient of velocity, and temperature to obtain a solution and check for sonic point conditions. If the solution does not pass through the sonic point, then we change the temperature supplied at $\rin$ and repeat the process until and unless
for a certain temperature at $\rin$ we obtain a $\vin=v_{\rm in}^\prime$ which on integration satisfies 
the sonic point conditions at some $r=r_c$. Therefore, we obtain a transonic solution by iterating the temperature
to give us the unique transonic solution. This is in essence a variation of the solution procedure of Becker and his collaborators.

For two-temperature flow however, we have three unknowns, $\vin$, $\theh$ and $\thph$ at $\rin$, and still
two constants of motion $E$ and $\dot{M}$. That is, the number of variables increases by one, while
the number of equations, or equivalently the number of constants, remains the same
as we had in the single temperature regime. So for a given $E$ and ${\dot M}$, we supply $\thphp$,
$\thehp$ at $\rin$ and compute $\vinp$ from the expression of $E$. Considering $\thphp$, $\thehp$ and $\vinp$
as guess values of temperatures and flow velocity near the horizon, we integrate equations (\ref{eq:dvdr}, \ref{eq:dtpdr} \& \ref{eq:dtedr}) outwards and check for sonic point conditions (equations \ref{eq:sp1}, \ref{eq:sp2}). If the sonic point condition is not satisfied, then we change the value of $\thehp$, obtain another value of $\vinp$ and again we integrate the same equations. Similarly we also change $\thphp$ and repeat the same
procedure again, if no transonic solution is obtained. If the sonic point is found out, then the transonic solution with those values of $\thph=\thphp$, $\theh=\thehp$ and $\vin=\vinp$ for that particular set of $E$ and ${\dot M}$, is the solution.
 We have chosen $\rin=2.001\rg$. One has to remember however, now the system is under determined, the consequence of which will be seen in the next section. It may be further noted that, we mentioned $\thph$ is supplied to iterate $\theh$ and $\vin$ from $E$, however while presenting results, we prefer to quote $\tph$ or $\teh$ instead. This will make it easier for the reader to relate to the problem.

\section{RESULT}
We initially assume $\qpp=0$ to discuss various features of two-temperature
solution. The effect of $\qpp \neq 0$ will be discussed later in section \ref{subsec:heating}. 

\subsection{Investigating degeneracy in two-temperature flows}
\label{subsec:degenret}

\begin{figure}
 \centering
 \includegraphics[width=1.0\columnwidth,trim= 5 300 0 0,clip]{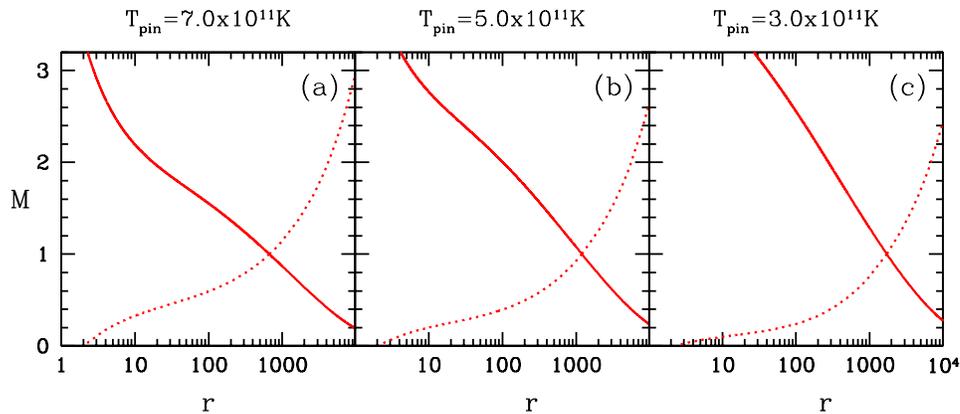}
 \caption{(a) Accretion $M$ (solid, red) and wind $M$ (dotted, red) as a function of $r$ corresponding to $M_{BH}=10 \msol$, $\dot{M}=0.01$ and $E=1.0001$. The different solutions are obtained changing $\tph$ (values are written on the top of each panel).}
 \label{fig:fig1}
\end{figure}

In Figs.(\ref{fig:fig1}a, b, c), we present the accretion solutions of two-temperature Bondi flow for $\mbh=10 \msol$, $\dot{M}=0.01$ and $E=1.0001$.
Each panel shows the accretion Mach number or $M=v/a$ (solid, red) and corresponding wind $M$ (dotted, red) as a function of $r$. The crossing points are the location of sonic/critical points.
The three solutions plotted in the figure are obtained by changing the proton temperature $\tph$
($\tph=\tp |_{r \rightarrow \rin}$), but for the same $E$ and ${\dot M}$ for a given central BH.
This implies that different values of $\tph$ would yield different solutions, each with a unique sonic point position and sonic point properties. 
In section \ref{subsec:method1}, we pointed out that the two-temperature regime is under determined, because we need to know
three unknowns at $\rin$ but there were only two constants of motion. The degeneracy in solution is the
direct fall out of such a scenario. All transonic two-temperature solutions, whether in exact GR or
in pseudo-Newtonian regime, suffers from this deficiency. 
In the next section we will discuss, the physical principle to be followed in order to obtain a unique
two-temperature transonic solution.

\begin{figure}
 \centering
\begin{centering}
 \includegraphics[width=1.0\columnwidth,trim= 0 0 0 0,clip]{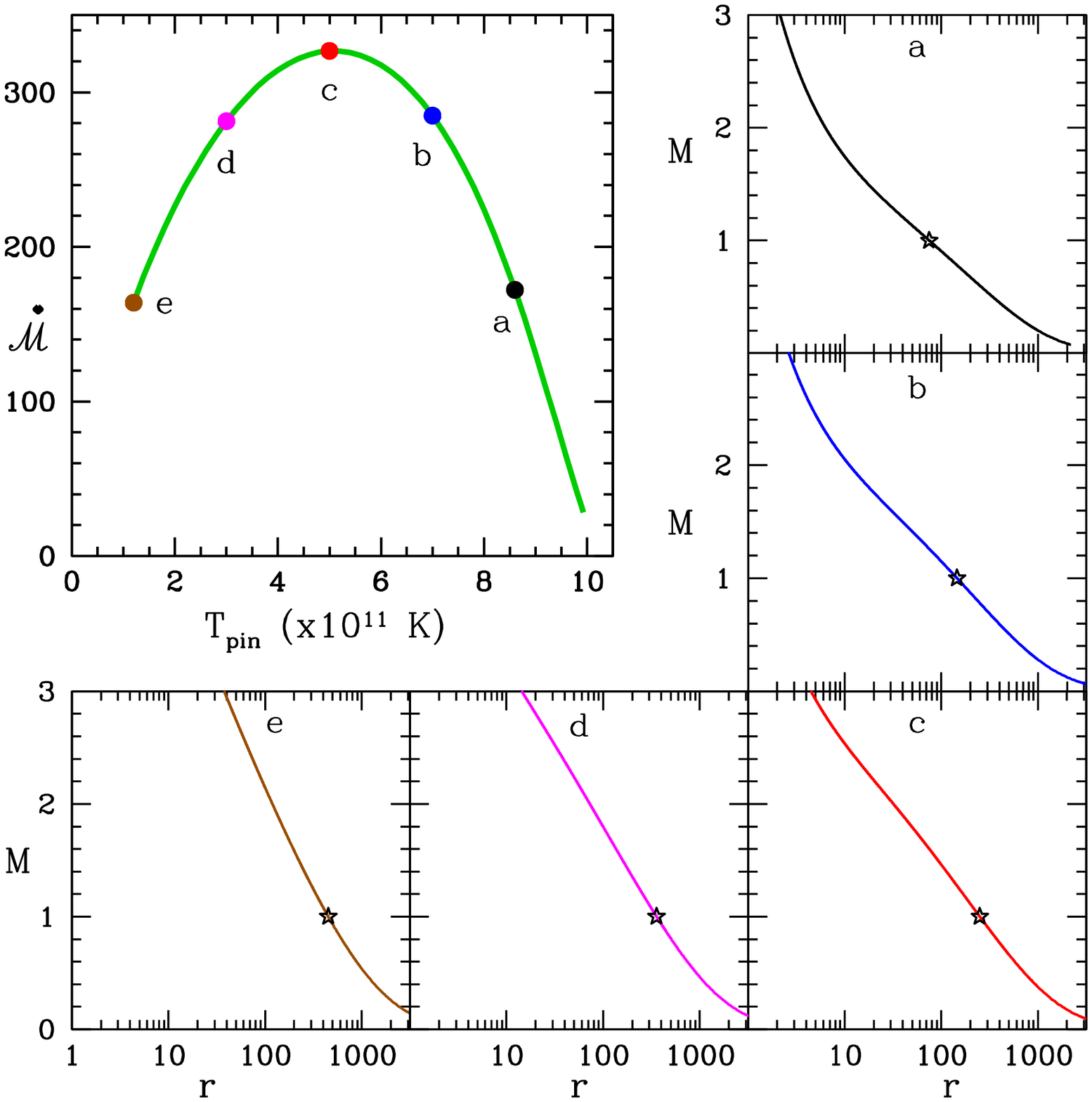}
 \caption{Top left panel: Variation of $\mdotin$ as a function of $\tph$ for accretion flow of $\dot{M}=0.1$ and $E=1.001$ onto a $10\msol$ BH. Panels `a' to `e' presents $M$ of the accretion (solid) with $r$ corresponding to each of the points `a'---`e' on the $\mdotin$--$\tph$ curve. The stars show the location of sonic points.
 At $\tph=5.0 \times 10^{11} K$ (marked `c') entropy maximizes, so panel `c' is the correct solution for the given $E$ and ${\dot M}$.}
 \label{fig:fig2}
\end{centering}
\end{figure}

\subsection{Entropy measure as a tool to remove degeneracy in two-temperature flows}
\label{subsec:uniqsoln}
As has been shown in Fig. (\ref{fig:fig1}a---c), for a given set of constants of motion namely $E$ and ${\dot M}$, there can be a plethora of transonic solutions, each differentiated by the $\tph$ at $\rin$.
Now the only way this degeneracy can be removed is by invoking the second law of thermodynamics.
It has also been shown in section \ref{subsec:entropy_accretion_formula}, that a general analytical expression
of entropy measure is not possible, however, the entropy of the accreting matter very close to the BH can be calculated (equation \ref{eq:ent_acc_rate}). So in Fig. (\ref{fig:fig2}, top left panel) we plot the measure of entropy
$\mdotin$ at $r=\rin$ as a function of $\tph$, for an accretion flow characterized by constants of motion
$\dot{M}=0.1$ and $E=1.001$ on to a BH of $M_{BH}=10\msol$.
We have marked points `a' to `e' on the $\mdotin$ vs $\tph$ curve, and then have plotted the
corresponding solutions ($M$ vs $r$) in the adjacent panels also named as `a' to `e'. 
It is easy to notice that the solutions are completely different since, the sonic points
of the solutions vary by a few $\times 100\rg$ for this particular $E$ and ${\dot M}$. In particular,
solution marked `a' and that marked `e' both have the same $\mdotin$ and $E$, but the sonic point of `a' is at
$r_c=75.008$ and that of `e' is at $r_c=451.297$, respectively. Different proportions of $\te$ and $\tp$
might give rise to the same $\mdotin$ and $E$! 
This also implies a wrong
choice of solution would lead us to wrong conclusions about the physical processes around BHs.
{\it However, only one of them is correct}.
It must be noticed that, of all the solutions, the entropy distribution has single well behaved maxima at $\tph=5\times 10^{11}$K, and therefore, by the second 
law of thermodynamics, the accretion solution corresponding to this entropy at point `c' on the curve
is the correct one.

\subsection{Properties of unique two-temperature transonic solution}
\subsubsection{Critical point properties:}
\label{subsubsec:crit}

\begin{figure}
 \centering
 \includegraphics[width=1.0\columnwidth,trim= 0 0 0 0,clip]{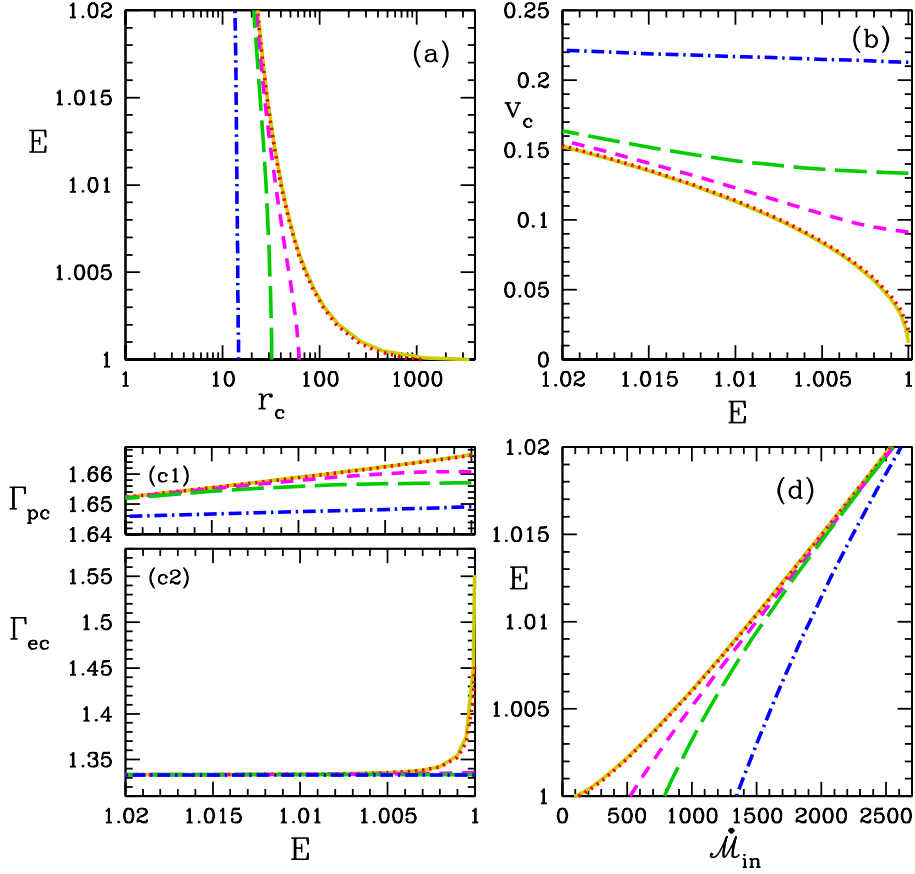}
 \caption{Variation of sonic points and its properties with the accretion rate ($\dot{M}$) of the BH. Here we have assumed $\mbh=~10\msol$. We have taken $\dot{M}=0.01$ (yellow-solid), $0.10$ (red-dotted), $0.50$ (magenta-dashed), $1.00$ (green - long-dashed) and $5.00$ (blue - dot-dashed).} 
 \label{fig:fig3}
\end{figure}

For adiabatic flow, sonic points can be found directly from a given value of $E$, but in our case the sonic point can be
obtained only after obtaining the solution. Since the system is under determined, unique $r_c$ can only be
obtained by invoking the second law of thermodynamics. Taking all these factors into consideration,
we plot $E$ as a function of $r_c$ (Figs. \ref{fig:fig3}a); while
$v_c$ (Fig. \ref{fig:fig3}b); $\gampc$, $\gamec$ (Fig. \ref{fig:fig3}c)
and $\mdotin$ (Fig. \ref{fig:fig3}d) as functions of $E$.
Each curves are for accretion rate $\dot{M}=0.01$ (yellow-solid), $0.10$ (red-dotted), $0.50$ (magenta-dashed), $1.00$ (green - long-dashed) and $5.00$ (blue - dot-dashed). Here a BH of $10\msol$ has been considered.
For low accretion rates (${\dot M}\leq 0.1$), the range of sonic points are $3<r_c \rightarrow \infty$, however,
for higher accretion rates, the sonic point range decreases significantly. In presence of significant cooling
(i. e., higher ${\dot M}$), hot flows from large distance can be accreted, which otherwise could not be accreted.
As a result $v_c$ and the entropy both are higher for flows with higher ${\dot M}$. From all the plots it is clear
that, for spherical accretion, there can be only one sonic point.

\begin{figure}
\begin{centering}
 \includegraphics[height=1.0\columnwidth,width=1.0\columnwidth,trim= 8 10 100 20,clip]{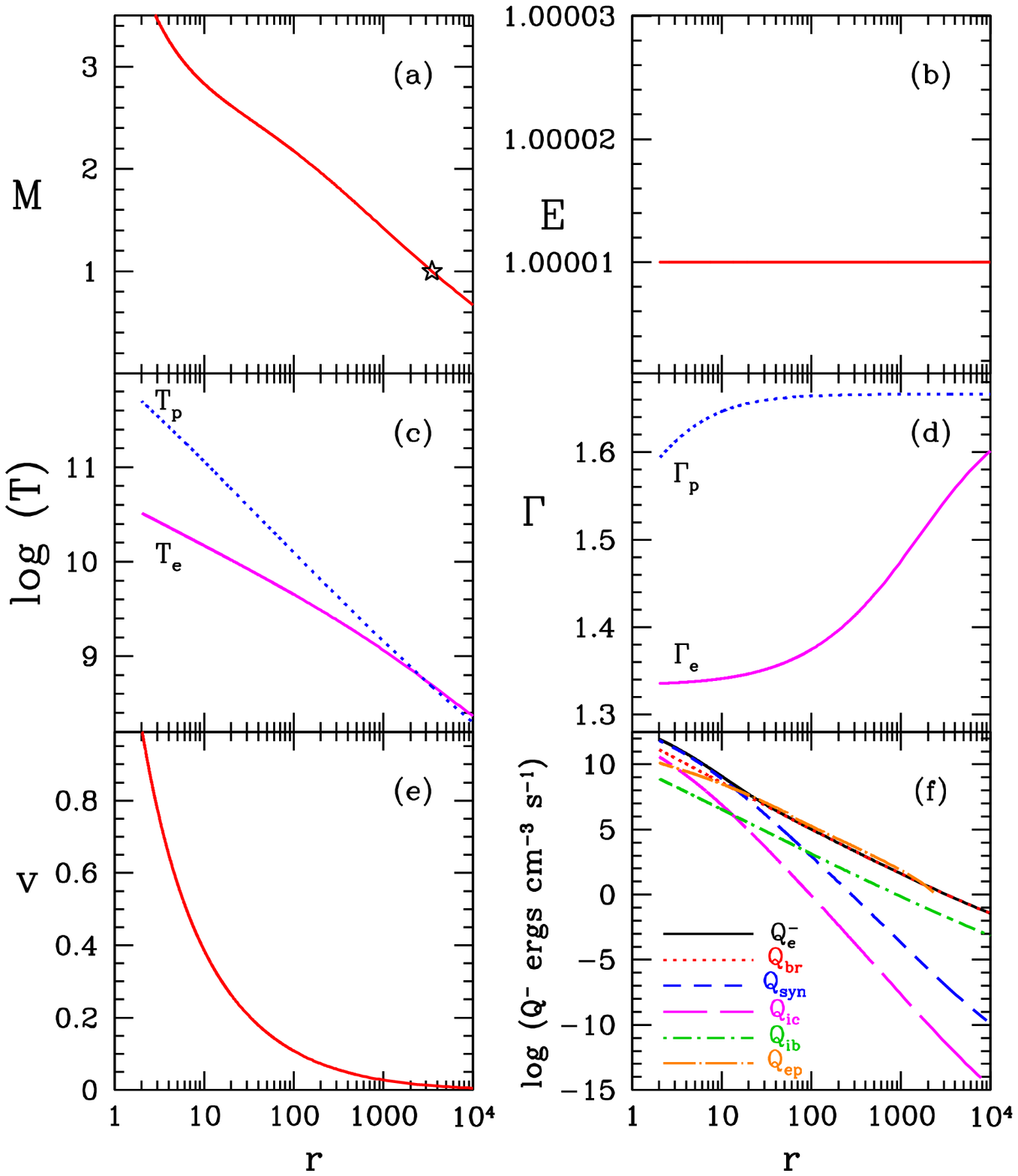}
 \caption{Variation of (a) $M$; (b) $E$; (c) $\tp$ and $\te$; (d) $\game$
 and $\gamp$; (e) $v$; and (f) total ($\qem$),  bremsstrahlung ($\qbr$), synchrotron ($\qsy$),  inverse-Compton ($\qic$), 
 and inverse-bremsstrahlung ($\qib$) emissivities as a function of $r$. The Coulomb coupling
 $\gamep$ is over plotted. The star on the $M$ distribution represent the
 location of the sonic point $r_c$. The accretion disc parameters are $E=1.00001$, $\mbh=10\msol$ and $\dot{M}=0.01$. The Qs presented, are in physical units (ergs cm$^{-3}$ s$^{-1}$).}
 \label{fig:fig4}
\end{centering}
\end{figure}

\begin{figure}
 \centering
\begin{centering}
 \includegraphics[width=1.01\columnwidth,trim= 0 0 0 0,clip]{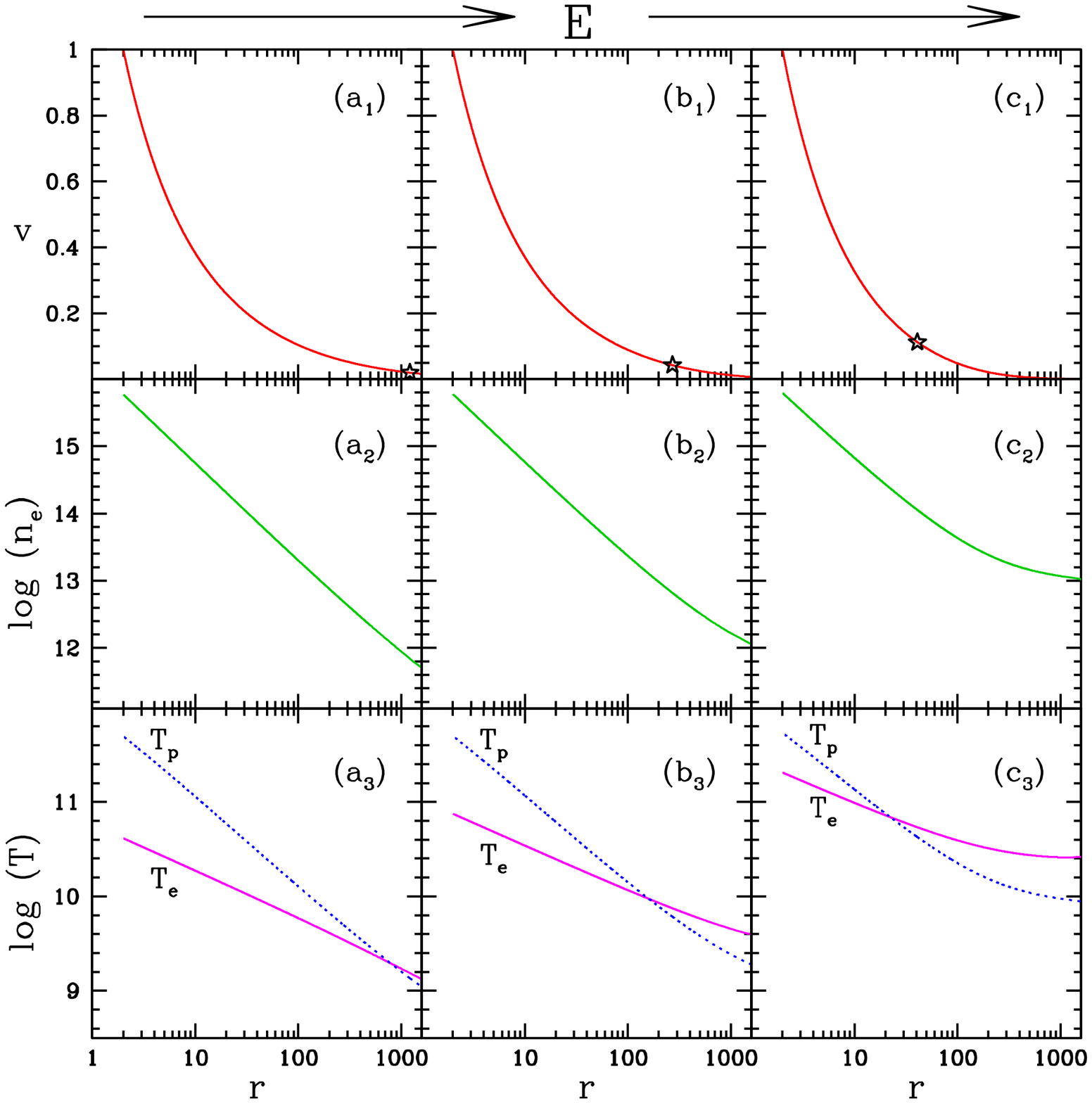}
 \caption{Variation of $v$ (a$_1$, b$_1$, c$_1$); electron number density $\nel$ (a$_2$, b$_2$, c$_2$)
 and $\tp$ \& $\te$ in panels (a$_3$, b$_3$, c$_3$), as a function of $r$.
The generalized Bernoulli parameter
changes from the left panels $E=1.0001$ (a$_1$, a$_2$, a$_3$), to the middle panels $E=1.001$ (b$_1$, b$_2$, b$_3$) and then to the right panels $E=1.01$ (c$_1$, c$_2$, c$_3$). 
Other parameters selected are $\mbh=10 \msol$ and ${\dot M}=0.01$.}
 \label{fig:fig5}
\end{centering}
\end{figure}

\subsubsection{Flow variables and emissivity}
In Fig.(\ref{fig:fig4}a-f) we present various flow variables of the correct Bondi accretion
on to a BH $\mbh=10\msol$.
The constants of motions are $E=1.00001$ and ${\dot M}=0.01$. The flow variables plotted are 
$M$, $E$, $\tp$ \& $\te$, $\game$ \& $\gamp$ and $v$ on the panels Fig.(\ref{fig:fig4}a-e), respectively. The star mark indicates the location of sonic point.
\textbf{Figure (\ref{fig:fig4}b) shows that} the generalized Bernoulli parameter $E$ is indeed a constant
of motion. It is also to be noted that, $\te\approx \tp$ (solid, Fig. \ref{fig:fig4}c) at large $r$ and
$\te < \tp$ at $2<r< 1000$. Moreover, the electron fluid while traveling a distance of about $10^4\rg$ on the way to the BH,
spans a temperature range of more than two orders of magnitude which means $1.6 > \game \sim 4/3$ and do not have any constant value. In addition,
$1.6 < \gamp \sim 5/3$ and the temperature of the proton fluid spans more than three orders of magnitude. But the distribution of $\game$ \& $\gamp$ would also change for a different
set of constants of motion ($E$, ${\dot M}$). In other words, considering CR EoS is important.
In Fig.(\ref{fig:fig4} f), we plot the total electron emissivity or $\qem$, bremsstrahlung ($\qbr$),
synchrotron ($\qsy$), inverse-Compton ($\qic$), inverse-bremsstrahlung ($\qib$) and Coulomb coupling term
($\gamep$) as a function of distance. All the Qs used are in physical units (ergs cm$^{-3}$ s$^{-1}$),
and for simplicity $\bm{\bar{Q}}$ are not used.
$\qbr$ dominates the radiative process for this particular set of $E$ and ${\dot M}$, except near the horizon
where
the $\qsy \gsim \qbr$. $\qic$ is quite weak for low
accretion rate. $\qib$ may have larger contribution than $\qsy$ or $\qic$ 
at larger distance, but $\qib \ll \qem$. Since $\gamep$ is also comparable to $\qem$ except near the horizon,
$\tp$ and $\te$ is comparable in a large range of $r$. Close to the horizon, $\gamep \ll \qem$ as a result
$\tp \gg \te$. On careful inspection it is clear that at $r>2000\rg$, $\gamep < 0$ and therefore $\te > \tp$.
So one can say, attainment of single temperature distribution or, two-temperature distribution depends on
the relative strength of Coulomb interaction and various radiative processes.

\subsubsection{Dependence of accretion flow on $E$ and ${\dot M}$}

\begin{figure}
 \centering
\begin{centering}
 \includegraphics[width=1.1\columnwidth,trim= 0 0 0 0,clip]{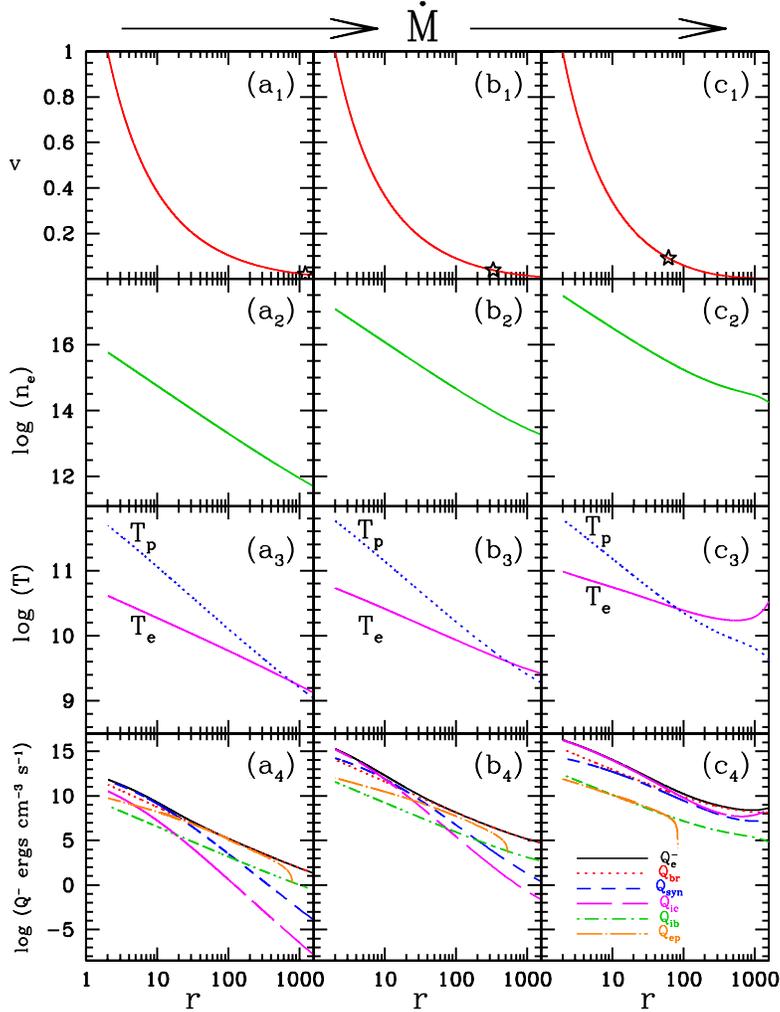}
 \caption{Variation of $v$ (a$_1$, b$_1$, c$_1$); electron number density $\nel$ (a$_2$, b$_2$, c$_2$),
$\tp$ \& $\te$ in panels (a$_3$, b$_3$, c$_3$) and various radiative emissivities and $\gamep$
(a$_4$, b$_4$, c$_4$)in panels  as a function of $r$.  
The accretion rate changes from the left panels ${\dot M}=0.01$ (a$_1$, a$_2$, a$_3$, a$_4$), to the middle panels
${\dot M}=0.2$ (b$_1$, b$_2$, b$_3$, b$_4$) and then to the right panels ${\dot M}=0.5$ (c$_1$, c$_2$, c$_3$,
c$_4$). 
Other parameters selected are $\mbh=10~\msol$ and $E=1.0001$. All Qs are presented in physical units (ergs cm$^{-3}$ s$^{-1}$).}
 \label{fig:fig6}
\end{centering}
\end{figure}

In Figs.(\ref{fig:fig5}a$_1$-c$_3$) we show how the global transonic two-temperature solutions
depend on constant of motion $E$ (increases left to right) for a given ${\dot M}$ on to a stellar mass BH.
We have plotted $v$ (a$_1$, b$_1$, c$_1$); electron number density $\nel$ (a$_2$, b$_2$, c$_2$)
 and $\tp$ \& $\te$ in panels (a$_3$, b$_3$, c$_3$), as a function of $r$.
The star on the velocity curve shows the location of sonic point. For higher $E$, $r_c$ is formed closer to the
horizon. Increasing $E$, raises the temperature at the outer boundary and reduces $v$, thus the electron number
density at the outer boundary is also higher for higher $E$.

Increasing ${\dot M}$ has similar effect on the accretion solutions. We plot the velocity distribution (Fig. \ref{fig:fig6}a$_1$, b$_1$, c$_1$),
$\nel$ (Fig. \ref{fig:fig6}a$_2$, b$_2$, c$_2$), $\te$ \& $\tp$ (Fig. \ref{fig:fig6}a$_3$, b$_3$, c$_3$)
and different radiative emissivities and $\gamep$ (Fig. \ref{fig:fig6}a$_4$, b$_4$, c$_4$)
as a function of $r$. Once again $Q$s presented in this figure are in physical units and we do not put
`bar' in order, not to make the figure clumsy. Keeping $E=1.0001$ constant, we change ${\dot M}=0.01$ (a$_1$ --- a$_4$), to 
${\dot M}=0.2$ (b$_1$ --- b$_4$) and then to ${\dot M}=0.5$ (c$_1$---c$_4$). 
Emission increases with the increase in
${\dot M}$, and therefore can accrete hotter flow at large distances. As a result the sonic points form closer to
the horizon, even for same $E$. The sonic point in the figure can also be seen to move closer to the horizon
(the star mark in the velocity distribution). For low ${\dot M}$, $\qbr$ dominates (see also Fig. \ref{fig:fig4} f). Interestingly, for a distance range of $20 < r < 1000$,
$\gamep \approx \qem$ (long dashed-dot).
Since Coulomb interaction is comparable to the bremsstrahlung emission, $\te \approx \tp$ in the same range.
As the
accretion rate increases, inverse-Compton cooling becomes more efficient and dominates in the overall emissivity
(Fig. \ref{fig:fig6}b$_4$).
The $\gamep$ term becomes less effective, as a result the difference between $\te$ and $\tp$ increases. For even higher 
${\dot M}$ (Fig. \ref{fig:fig6}c$_4$), inverse-Compton dominates the cooling and Coulomb term becomes even weaker and therefore $\te$ and $\tp$ becomes significantly different from each other. In fact, Coulomb coupling is effective when emission process is not very strong.

\begin{figure}
 \centering
\begin{centering}
 \includegraphics[width=1.01\columnwidth,trim= 0 140 0 0,clip]{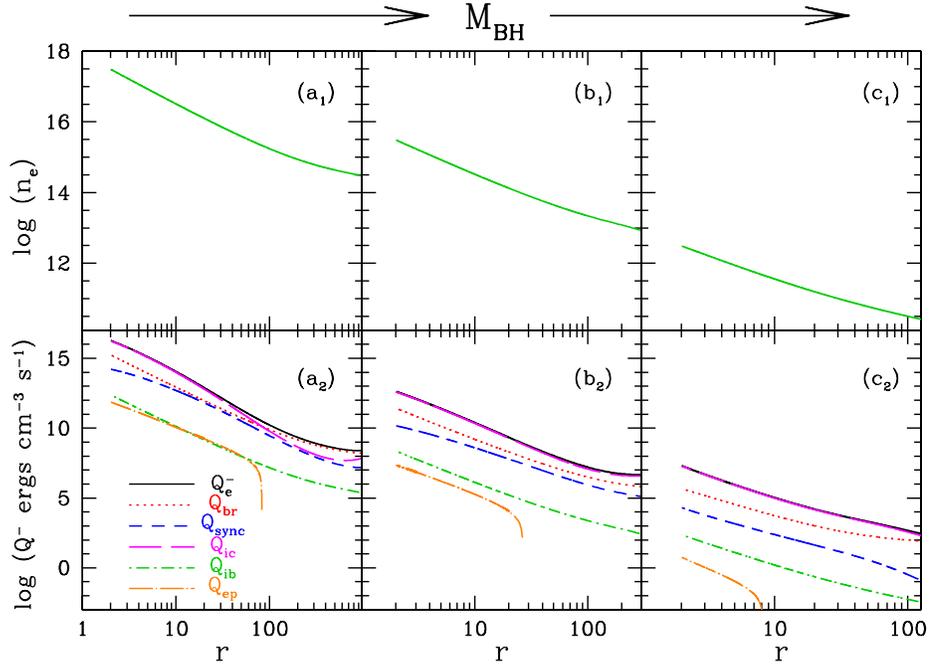}
 \caption{Variation of $\nel$ (a$_1$, b$_1$, c$_1$); emissivities and Coulomb coupling (a$_2$, b$_2$, c$_2$)
as function of $r$. Left column panels (a$_1$ and a$_2$) are for $\mbh=10 \msol$, the
middle column are for $\mbh=10^3 \msol$ (b$_1$, b$_2$)
and for right column $\mbh=10^6 \msol$ (c$_1$, c$_2$). 
Other parameters selected are $E=1.0001$ and ${\dot M}=0.5$. The Qs are in physical units (ergs cm$^{-3}$ s$^{-1}$).}
 \label{fig:fig7}
\end{centering}
\end{figure}

\subsubsection{Effect of the mass of the central BH}

Since the mass supplied is described in the units of Eddington rate, so the net amount of mass flux
increases with the central mass of the BH. The number density is proportional to the inverse of $\mbh$,
but the volume would increase as $\mbh^3$. Therefore, emissivity is proportional to $\mbh^{-2}$. 
As a result net radiative cooling increases with $\mbh$. This allows hotter matter to flow onto a more massive BH,
which pushes the sonic point closer to horizon even for matter starting with same $E$ and ${\dot M}$ (in units of Eddington rate).
 We plot $\nel$ (Fig. \ref{fig:fig7}a$_1$, b$_1$, c$_1$) and $\qem,~\qbr,~\qsy,~\qic,~\qib,~\gamep$ (Fig. \ref{fig:fig7}a$_2$, b$_2$, c$_2$)
 as a function of $r$, but for different $\mbh=10 \msol$ (Fig. \ref{fig:fig7}a$_1$, a$_2$), $\mbh=10^3 \msol$
 (Fig. \ref{fig:fig7}b$_1$, b$_2$) and $\mbh=10^6 \msol$ (Fig. \ref{fig:fig7}c$_1$, c$_2$).
 The Qs are presented in physical units.
 The sonic point for $\mbh=10 \msol$ is at $r_c=61.535$, for $\mbh=10^3 \msol$ the $r_c=39.966$ and finally for $\mbh=10^6 \msol$ the sonic point is at $r_c=19.786$.
 So it is clear that radial accretion onto larger BH, is hotter and will be more luminous than the smaller ones.
 For low accretion rates where the number density is lower, $\qic$ is generally lower than $\qbr$ or $\qsy$.
 But for higher ${\dot M}$ accretion, $\qic$ starts to dominate in the inner region. And since
 accreting larger BHs are more luminous, the total emissivity is dominated by $\qic$.
 These plots also shows that, for lower mass BH and higher ${\dot M}$, $\qsy$ is similar to $\qbr$, however,
 for higher $\mbh$, $\qbr$ is much stronger than $\qsy$. Whatever may be the mass of the central BH
 or accretion rate, $\qib$ is significantly lower than the net emissivity. The Coulomb coupling term $\gamep$
 is negligible for high ${\dot M}$ and decreases even more for flow around massive BHs. 
 
\subsection{Luminosity and efficiency of the systems}
\begin{figure}
	\centering
	\begin{centering}
		\includegraphics[width=1.0\columnwidth,trim= 0 260 0 0,clip]{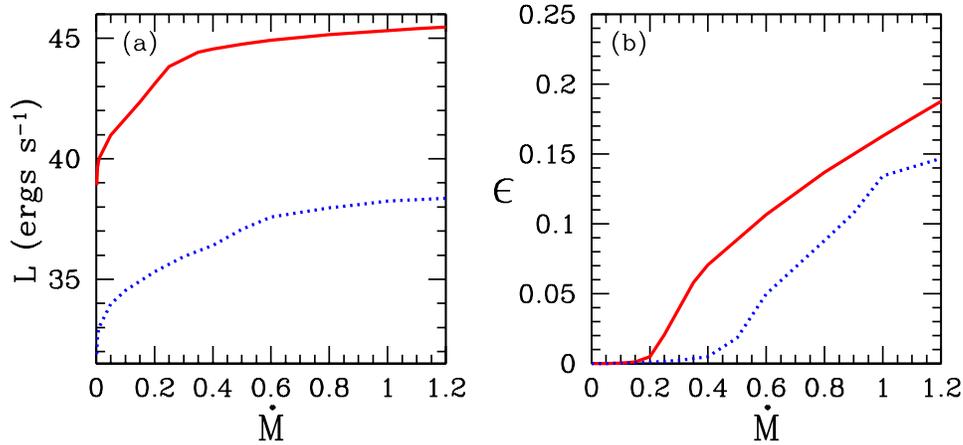}
		\caption{(a) Luminosity $L$, and (b) efficiency $\epsilon$ as a function of $\dot{M}$.
Each curve corresponds to $\mbh=10^8\msol$ (solid, red) and $\mbh=10 \msol$ (dotted, blue).
Other parameter is $E=1.001$.}
		\label{fig:fig8}
	\end{centering}
\end{figure}

Shapiro (1973)\cite{shapiro73} computed luminosity from Bondi flow via only the bremsstrahlung process,
and concluded that radial flow is not efficient enough.
However, that accretion model was not strictly two-temperature. Moreover, all classes of solutions were not
investigated. From Figs. (\ref{fig:fig4} --- \ref{fig:fig6}) of this paper, it is quite clear that the different cooling processes
start to dominate at different ${\dot M}$. For lower ${\dot M}$ inverse-Compton is not a very dominant process,
while for higher accretion rate, inverse-Compton becomes important. Therefore, it can be safely assumed that
both luminosity and efficiency of the accretion flow would also depend on the accretion rates. 

In Fig. {\ref{fig:fig8}, panel (a), we plot the variation in luminosity ($L$) in units of ergs s$^{-1}$, with $\dot{M}$ for accretion flow on to $\mbh=10 \msol$ (dotted, blue),
and $\mbh=10^8 \msol$ (solid, red). Other parameter of the flow is $E=1.001$. The efficiency of a BH system can be written as $\epsilon=L/(\dot{M}c^2)$. In Fig. (\ref{fig:fig8} b), the corresponding $\epsilon$ is plotted as a function of $\dot{M}$. For low ${\dot M}\lesssim 0.2$, the efficiency of conversion of accretion energy to
radiation is really low $\epsilon \lesssim 0.01$ for both kind of BHs. However, for ${\dot M}>0.5$ the efficiency $\gtrsim 0.1$ for accretion on to $10^8 \msol$ BH and comfortably produces $L\gtrsim 10^{44}$ergs s$^{-1}$. At super Eddington accretion rates super massive BH produces luminosities above $10^{45}$ erg s$^{-1}$ with efficiency
$\epsilon \sim 0.2$.
Accretion flow on to stellar mass BH can emit at $L \sim 10^{38}$erg s$^{-1}$ for ${\dot M}\gtrsim 0.8$.
However, the efficiency of accretion flow around stellar mass BH is generally lower than the one around super-massive BH, however, for ${\dot M}>0.8$ the efficiency $\epsilon > 0.1$. So not only the accretion flow onto massive BHs are brighter, even its radiative efficiency is more. 

\begin{figure}
\centering
\begin{centering}
\includegraphics[width=1.0\columnwidth,trim= 0 0 100 0,clip]{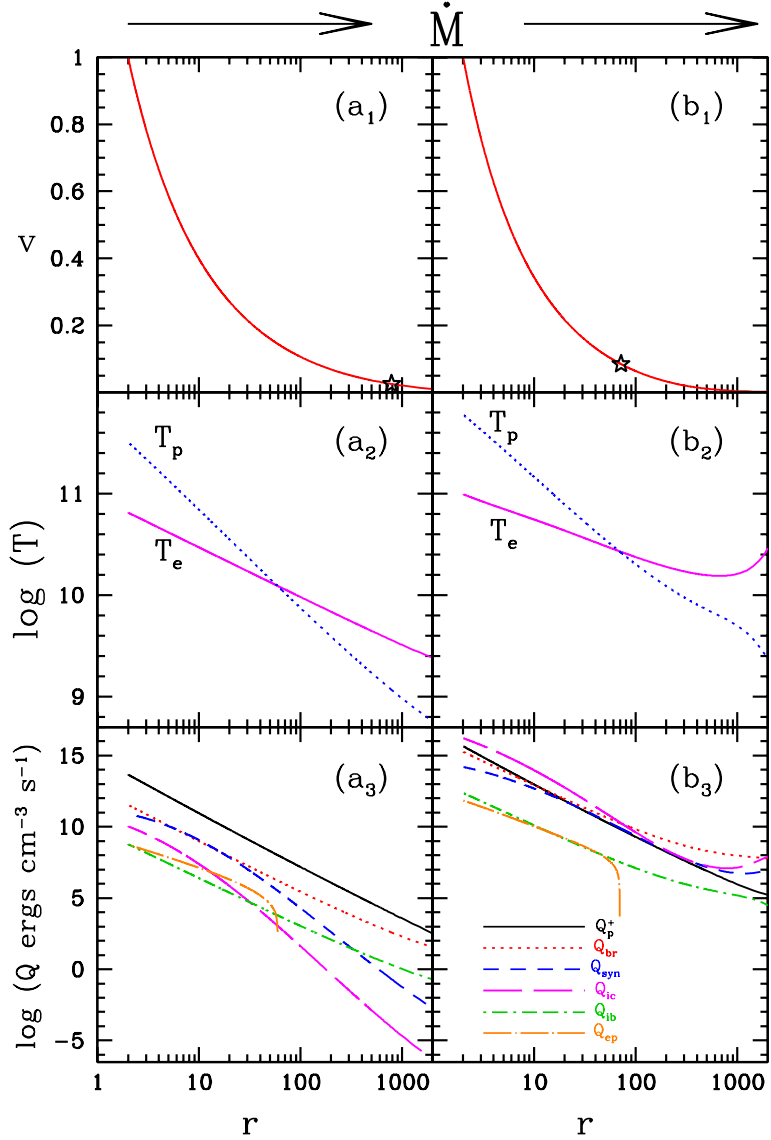}
\caption{Three-velocity $v$ (a$_1$, b$_1$), temperatures (a$_2$, b$_2$) and emissivities, heating and Coulomb coupling (a$_3$, b$_3$) as a function of $r$. The solutions are for ${\dot M}=0.01$ (a$_1$---a$_3$)
and ${\dot M}=0.5$ (b$_1$---b$_3$)
Other parameters are for $E=1.001$ and $\mbh=10\msol$.}
\label{fig:fig9}
\end{centering}
\end{figure}

\subsection{Effect of dissipative proton heating}
\label{subsec:heating}
So far in this paper we considered no explicit heating. We now consider dissipative magnetic heating 
in the footsteps of Ipser \& Price (1982)\cite{ip82} . It mainly affects the protons, however, through
Coulomb coupling the dissipated energy is also transmitted to the electrons.
In Fig. (\ref{fig:fig9}a$_1$---a$_3$), we plot $v$ (panel a$_1$), temperatures (panel a$_2$) and
various emissivities, heating rate and the Coulomb coupling term (panel a$_3$). Comparing with Fig. (\ref{fig:fig5}b$_1$---b$_3$), which was for the same accretion parameters but without heating, the effect of heating is clearly seen.
The sonic point in the present case is pushed back, i.e., BH is accreting matter with lower temperatures
at the outer boundary. Since the ${\dot M}$ is low, so the heating term $\qpp$ dominates.
In Fig. (\ref{fig:fig9}b$_1$---b$_3$) the same variables are plotted but now for higher ${\dot M}=0.5$.
In this case the $\qem$ dominates over $\qpp$. The Coulomb coupling on either case is negligible.
Heating processes quantitatively affects the solutions, if the dissipative heat only directly affects the protons.
This is because, in general Coulomb coupling is not very effective in energy exchange between electrons and protons and was also suggested by Manmoto et. al (1997)\cite{manmotoetal97}.

\section{DISCUSSIONS AND CONCLUSIONS}

A correct two-temperature solution is very important, because a proper electron temperature distribution for a given boundary condition, produces the correct spectrum and luminosity. 
Moreover, analytical solutions obtained in this paper is also important since, these solutions may act
as tests, as well as, may be used as initial conditions for simulation codes.

Although there are few papers in the single temperature
regime, which used constants of motion to obtain the solutions, but as far as we know, probably there are none in the two-temperature domain which even addresses the issue of constants of motion while obtaining the solutions.
It may be remembered that, a fluid solution is not just characterized by its energy but also its entropy,
and according to the second law of thermodynamics, any physical solution should correspond to the one with highest entropy.
It was Bondi\cite{bondi1952} who used this principle in order to stress that a transonic solution is the
correct accretion solution under the influence of gravity. Later Becker and his collaborators\cite{bl03,bdl08,lwwbp16,
lb17} used the information of energy as well as the entropy to obtain transonic accretion solutions around
a black hole in presence of dissipation. Since the set of equations in single temperature flow is complete,
so finding a transonic solution suffices the criteria for second law of thermodynamics.
However, as has been discussed extensively in the paper, the set of governing equations are less than the number of variables, second law of thermodynamics
becomes essential even to find a proper solution. The novelty of this work is to identify this problem and
laying down the procedure to overcome it, by actually following the footsteps of Bondi and Becker.

In this paper, we obtained the expression for the generalized Bernoulli parameter ($E$) for two temperature
flow, by integrating
the energy-momentum balance equation and showed that it is indeed a constant of motion. Moreover,
integrating the continuity equation we obtained the expression for accretion rate (${\dot M}$) which is the other
constant of motion.
In addition, we explicitly showed that degenerate transonic solutions exist for a given set of constants of motion. To remove the degeneracy we took the help of the second law of thermodynamics
near the horizon, according to which the transonic solution which has maximum entropy should be the
solution. The next hurdle was, that there was no analytical expression of entropy measure for two-temperature
flow. We used the BH inner boundary condition (gravity overwhelms all other interactions), in order
to obtain the analytical expression of entropy measure ($\mdotin$) for a gas in two-temperature regime valid
only near the horizon and that too, by using relativistic EoS.

To focus on the problem of degenerate two-temperature solutions and its possible remedy, we considered
a simple accretion model of radial flow onto a black hole. More complicated accretion model would have obscured the crux of the problem. Simple as it may be, but spherical accretion is preferred mode of accretion onto isolated BHs immersed in interstellar matter and has been
shown by many authors\cite{dp80,bk05} . Moreover, the inner region of a BH accretion disc is also quasi spherical
and many researchers have considered radial inflow to mimic inner accretion disc\cite{tmk97,kht97} . Since
radial flow has no angular momentum (quasi spherical flow may have minuscule amount), viscous transport
should be negligible for accretion onto isolated BH or in the inner region of an accretion disc.
Moreover, authors who have obtained transonic two-temperature solution before\cite{manmotoetal97} , are of the view that Coulomb coupling is not an efficient energy transfer process. Therefore any viscous heating will anyway
not find its way into heating up the electrons. Looking into all these factors, we ignored viscous dissipation
and concentrated two-temperature accretion flow by only considering cooling mechanisms in this paper.
However, at the end we did consider dissipative proton heating\cite{ip82} . Heating has quantitative effect
on the accretion solutions
and confirmed that Coulomb coupling is weak as was mentioned by Manmoto et. al. (1997)\cite{manmotoetal97} . 

Using the methodology explained above, we obtained all possible solutions for any given set of $E$ and ${\dot M}$.
For higher $E$ and higher ${\dot M}$, sonic points were formed closer to the horizon, while for lower values of
both the constants of motion, sonic points occurred at larger distances. We showed that for correct solutions
the adiabatic index of electron and proton fluid varies from non-relativistic to relativistic values.
We also showed that, different cooling processes become important for different values of ${\dot M}$.
Therefore radiative efficiency depends on ${\dot M}$. For ${\dot M}<0.1$, whether it is a super massive BH or a
stellar one, the accretion flow is inefficient. However for super-massive BH, the accretion flow becomes radiatively efficient i.e., more than $10\%$ for ${\dot M}\gtrsim 0.6$. For stellar mass BH, the accretion becomes
radiatively efficient when the accretion rate is close to Eddington rate. It is observed that whenever
local inverse-Compton processes dominate, the accretion flow becomes luminous. Therefore, it is not
necessary that radial accretion is radiatively inefficient. 


\end{document}